\documentclass[reprint, amssymb, amsmath, prb,cha]{revtex4-1}
\usepackage{graphicx}
\usepackage{bm}
\usepackage{soul}

\usepackage[colorlinks=true,linkcolor=blue]{hyperref}
\expandafter\ifx\csname package@font\endcsname\relax\else
 \expandafter\expandafter
 \expandafter\usepackage
 \expandafter\expandafter
 \expandafter{\csname package@font\endcsname}
\fi
\usepackage{color}

\begin{document}

\title{Effective $n$-type Doping of Mg$_3$Sb$_2$ with Group-3 Elements}

\author{Prashun Gorai}
\email{pgorai@mines.edu}
\author{Eric S. Toberer}
\author{Vladan Stevanovi\'c}
\email{vstevano@mines.edu}
\affiliation{Colorado School of Mines, Golden, Colorado 80401, USA}%

\begin{abstract}
The recent discovery of high thermoelectric performance in Mg$_3$Sb$_2$ has been critically enabled by the success in $n$-type doping of this material, which is achieved under Mg-rich growth conditions, typically with chalcogens (Se, Te) as extrinsic dopants. Using first-principles defect calculations, we previously predicted that higher electron concentrations ($\sim10^{20}$ cm$^{-3}$) can be achieved in Mg$_3$Sb$_2$ by doping with La instead of Se or Te.\cite{Gorai2018} Subsequent experiments\cite{imasato2018} showed that free electron concentration in La-doped Mg$_3$Sb$_{2-x}$Bi$_x$ indeed exceeds those in the Te-doped material. Herein, we further investigate $n$-type doping of Mg$_3$Sb$_2$ and predict that, in addition to La, other group-3 elements (Sc, Y) are also effective as $n$-type dopants; Y is as good as La while Sc slightly less. Overall, we find that doping with any group-3 elements should lead to higher free electron concentrations than doping with chalcogens.
\end{abstract}
\maketitle

\section{Introduction}
In thermoelectric materials, the concentration of free charge carriers needs to be highly tunable to allow optimization of the thermoelectric figure of merit \textit{zT}.\cite{toberer2008,gorainrm2017} Therefore, the ability to dope a material to the desired charge carrier type ($n$ or $p$) and carrier concentration plays a vital role. Beyond thermoelectrics, dopability of semiconductors is critical for optoelectronic\cite{nakamura1994,hosono2017} and power electronic applications.\cite{hudgins2003} 

Mg$_3$Sb$_2$-based thermoelectrics have recently risen to prominence almost exclusively due to the success in $n$-type doping that was key to achieving high \textit{zT}.\cite{tamaki2016,ohno2017,zhang2017exp} Additionally, the earth-abundance and non-toxicity of the constituent elements make Mg$_3$Sb$_2$ a very attractive material for thermoelectric applications. The $n$-type doping of Mg$_3$Sb$_2$, which has been achieved only recently\cite{tamaki2016,ohno2017,zhang2017exp}, requires Mg-rich conditions during growth to suppress the formation of compensating, acceptor-behaving Mg vacancies. In addition, one needs to introduce extrinsic dopants, most commonly chalcogens such as Se and  Te, which behave as donor defects when substituting Sb. Combined with Bi alloying to reduce the lattice thermal conductivity, a high \textit{zT}\textgreater1.5 has been demonstrated in Te-doped Mg$_3$Sb$_2$-based materials.\cite{ohno2017,imasato2018} 

Using first-principles defect calculations, we have previously examined the role of native defects and growth conditions in the dopability of Mg$_3$Sb$_2$.\cite{ohno2017,Gorai2018} We found that under Mg-poor conditions, the predominant defects are acceptor Mg vacancies that inhibit $n$-type doping, consistent with the observations in previous works\cite{tamaki2016}. In contrast, under Mg-rich conditions, donor-behaving Mg interstitials are the predominant defects. As a result of the native defect energetics, we found that Mg$_3$Sb$_2$ is natively self-doped $p$-type under Mg-poor and $n$-type under Mg-rich growth conditions. More importantly, the low concentrations of acceptor Mg vacancies under Mg-rich conditions allows extrinsic $n$-type doping with Te\cite{ohno2017,imasato2018}, with predicted free electron concentrations exceeding 10$^{19}$ cm$^{-3}$ at synthesis temperatures of $\sim$900 K, consistent with the experimental findings\cite{ohno2017}.

\begin{figure}[t]
\centering
\includegraphics[width=0.9\linewidth]{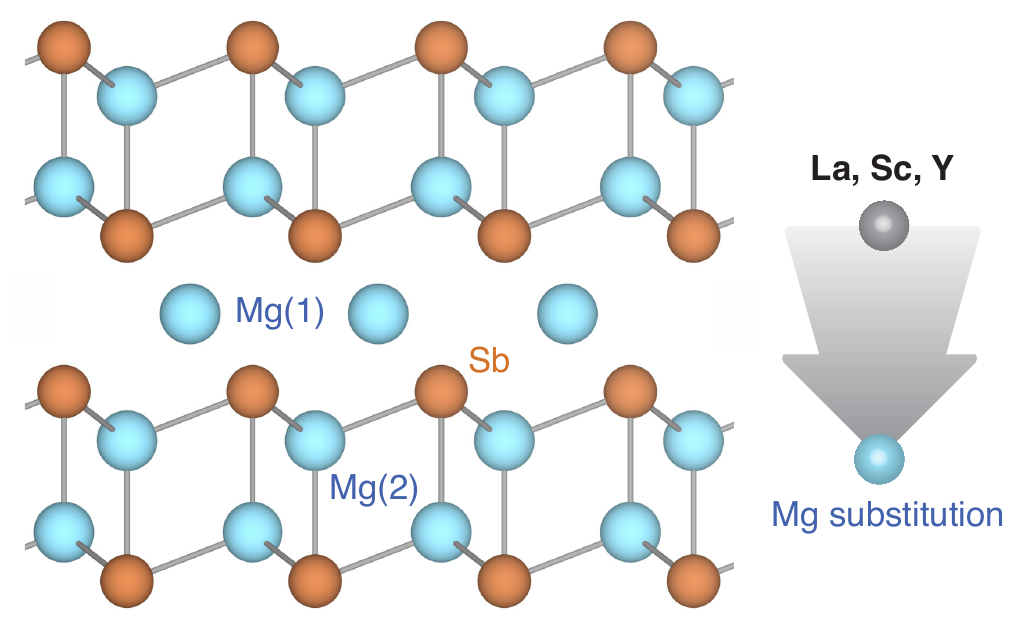}
\caption{\label{structure}
Mg$_3$Sb$_2$ contains two unique Mg Wyckoff positions denoted by Mg(1) and Mg(2), and one unique Sb Wyckoff position. The crystal structure of Mg$_3$Sb$_2$ can be visualized as slabs of [Mg$_2$Sb$_2$]$^{2-}$ intercalated with Mg$^{2+}$ cations. An effective $n$-type doping strategy involve Mg substitution with tri-valent group-3 elements (La, Sc, Y).}
\end{figure}

In addition to Sb substitution with Se and Te, we have previously investigated\cite{Gorai2018} other $n$-type doping strategies for Mg$_3$Sb$_2$, including: (a) Sb substitution with halides Br and I, (b) Mg substitution with trivalent cations La, Al, Nb, and Ga, and (c) insertion of cation interstitials Li, Zn, Cu, and Be. We found La to be the most effective $n$-type dopant of all the studied elements, and Mg substitution the most effective doping strategy. According to our calculations, La preferentially substitutes the Mg atoms (labelled Mg(1)) located between the [Mg$_2$Sb$_2$]$^{2-}$ layers, as shown in Figure \ref{structure}. The predicted electron concentration in La-doped Mg$_3$Sb$_2$ is $\sim$5$\times$10$^{20}$ cm$^{-3}$ (at 900K), which is almost an order of magnitude higher than for Te doping. This theoretical prediction has been experimentally validated;\cite{imasato2018} La is found to be an effective $n$-type dopant in Mg$_3$Sb$_{1.5}$Bi$_{0.5}$ with free electron concentrations nearly twice of those achieved through Te doping. Addition of La is shown to also significantly improve the thermal stability of $n$-type Mg$_3$Sb$_{1.5}$Bi$_{0.5}$. \cite{imasato2018} 

In this work, we extend our investigation of extrinsic $n$-type doping of Mg$_3$Sb$_2$ to other group-3 elements such as Sc and Y. We predict that, in addition to La, both Sc and Y are effective $n$-type dopants. We find that both Sc and Y are more effective than Te in terms of the free electron concentrations, which are predicted to exceed 10$^{20}$ cm$^{-3}$ (900 K); Y is as effective as La while Sc only slightly lesser. Overall, we find that out of the 15 different extrinsic dopants we have studied, Mg substitution with group-3 elements (Sc, Y, La) is by far the most effective $n$-type doping strategy. 

\section{Computational Methods}
First-principles point defect calculations provide formation energies of native defects and extrinsic dopants as functions of the Fermi energy. The achievable dopant and charge carrier concentrations in semiconductors can be predicted by combining the defect formation energetics with thermodynamic modeling of defect and charge carrier equilibria \cite{Goyal2017} We calculated the defect formation energies in Mg$_3$Sb$_2$ using density functional theory (DFT) and the standard supercell approach.\cite{lany2008} Within the supercell approach, the formation energy ($\Delta H_{D,q}$) of a point defect $D$ in charge state $q$ is calculated as:
\begin{equation}\label{defectenthalpy}
\Delta H_{D,q} = (E_{D,q} - E_H) + \sum_{i} n_i\mu_i + qE_F + E_{corr}
\end{equation}
where $E_{H}$ and $E_{D,q}$ represent the total energies of the defect-free, host supercell with no net charge ($E_H$) and the supercell with defect $D$ in charge state $q$, respectively. The chemical potential of element $i$ is denoted by $\mu_i$ and $n_i$ is the number of atoms of element $i$ added ($n_i$\textless 0) or removed ($n_i$\textgreater 0) from the supercell. $E_F$ is the Fermi energy, which represents the reservoir of charge (the Fermi sea). The term $qE_F$ in Eq. \ref{defectenthalpy} is the characteristic energy of exchanging charge between the defect and the reservoir of charge. The supercell approach to calculating defect energetics suffers from artifacts arising due to finite size effects. Additional artifacts are introduced due to the limitations of DFT, most notably, the underestimation of the band gap with standard functionals such as GGA-PBE.\cite{Perdew1996} Various correction schemes are available to correct for the finite size and insufficiently accurate electronic structure artifacts; these corrections, which are briefly discussed in the following paragraphs, are represented by the term $E_{corr}$. Detailed descriptions of the correction scheme can be found in Ref.~\citenum{lany2008}.

The total energies of the supercells we calculate using the generalized gradient approximation of Perdew-Burke-Ernzerhof (PBE) \cite{Perdew1996} within the projector augmented wave (PAW) formalism as implemented in the VASP software package.\cite{Kresse1996} A 3$\times$3$\times$2 supercell containing 90 atoms is used to calculate the defect energetics. The total energies are calculated with a plane-wave energy cutoff of 340 eV and a $\Gamma$-centered 4$\times$4$\times$4 Monkhorst pack $k$-point grid to sample the Brillouin zone. The defect supercells are relaxed following a similar procedure used in Refs. \citenum{ortiz2017,Gorai2018,ohno2017}. Specifically, 

The elemental chemical potentials $\mu_i$ are expressed relative to those of the elements in reference elemental phases as $\mu_i=\mu_i^0 + \Delta\mu_i$, where $\mu_i^0$ is the reference chemical potential under standard conditions and $\Delta\mu_i$ is the deviation from the reference. $\Delta\mu_i=0$ corresponds to $i$-rich conditions. For example, $\Delta\mu_\mathrm{Mg}=0$ corresponds to the equilibrium between Mg$_3$Sb$_2$ and Mg-metal conditions (referred to as Mg-rich). Following the FERE approach,\cite{stevanovic2012} the reference chemical potentials ($\mu_i^0$) are fitted to a set of measured formation enthalpies of compounds. The stability Mg$_3$Sb$_2$ relative to decomposition into competing phases determines the region of phase stability and the bounds on the values of $\Delta\mu_i$.

The underestimation of the band gap in DFT is remedied by applying individual valence and conduction band edge shifts (relative to the DFT-computed band edges) as determined from GW quasi-particle energy calculations \cite{lany2008}. We use DFT wave functions as input to the GW calculations. The GW eigen-energies are iterated to self-consistency to remove the dependence of the G$_0$W$_0$ result on the single-particle energies of the initial DFT calculation. The input DFT wave functions are kept constant during the GW calculations, which allows the interpretation of the GW quasi-particle energies in terms of energy shifts relative to the DFT Kohn-Sham energies. Relative to DFT-calculated conduction and valence band edges, the shifts are +0.176 eV (shift to higher energy) and -0.171 eV (shift to lower energy), respectively. 

The finite-size corrections that are included in $E_{corr}$, following the methodology in Ref. \citenum{lany2008}, are: (1) image charge correction for charge defects, (2) potential alignment correction for charged defects, (3) band filling correction for shallow defects, and (4) correction of band edges for shallow acceptors/donors. The calculations are organized and the results are analyzed using our software package, pylada-defects, for automation of point defect calculations.\cite{Goyal2016}

Under a given growth condition, the equilibrium $E_F$ and the corresponding free charge carrier concentration is determined by solving charge neutrality. The concentration of donor and acceptor defects are determined using Boltzmann distribution, such that $\left[D_q\right] = N_{s} e^{-\Delta H_{D,q}/\mathrm{k_B}T}$, where $\left[D_q\right]$ is the defect concentration, $N_s$ is the concentration of lattice sites where the defect can be formed, $\mathrm{k_B}$ is the Boltzmann constant, and T is the temperature. At a given T, The concentrations of electrons and holes are functions of $E_F$. To establish charge neutrality, the total positive charges should equal the negative charges. In this equation, $E_F$ is the only free parameter. By solving charge neutrality condition self-consitently, we can determine the equilibrium $E_F$ and the corresponding free carrier concentration.

\section{Results and Discussion}
%
\begin{figure}[!t]
\centering
\includegraphics[width=0.9\linewidth]{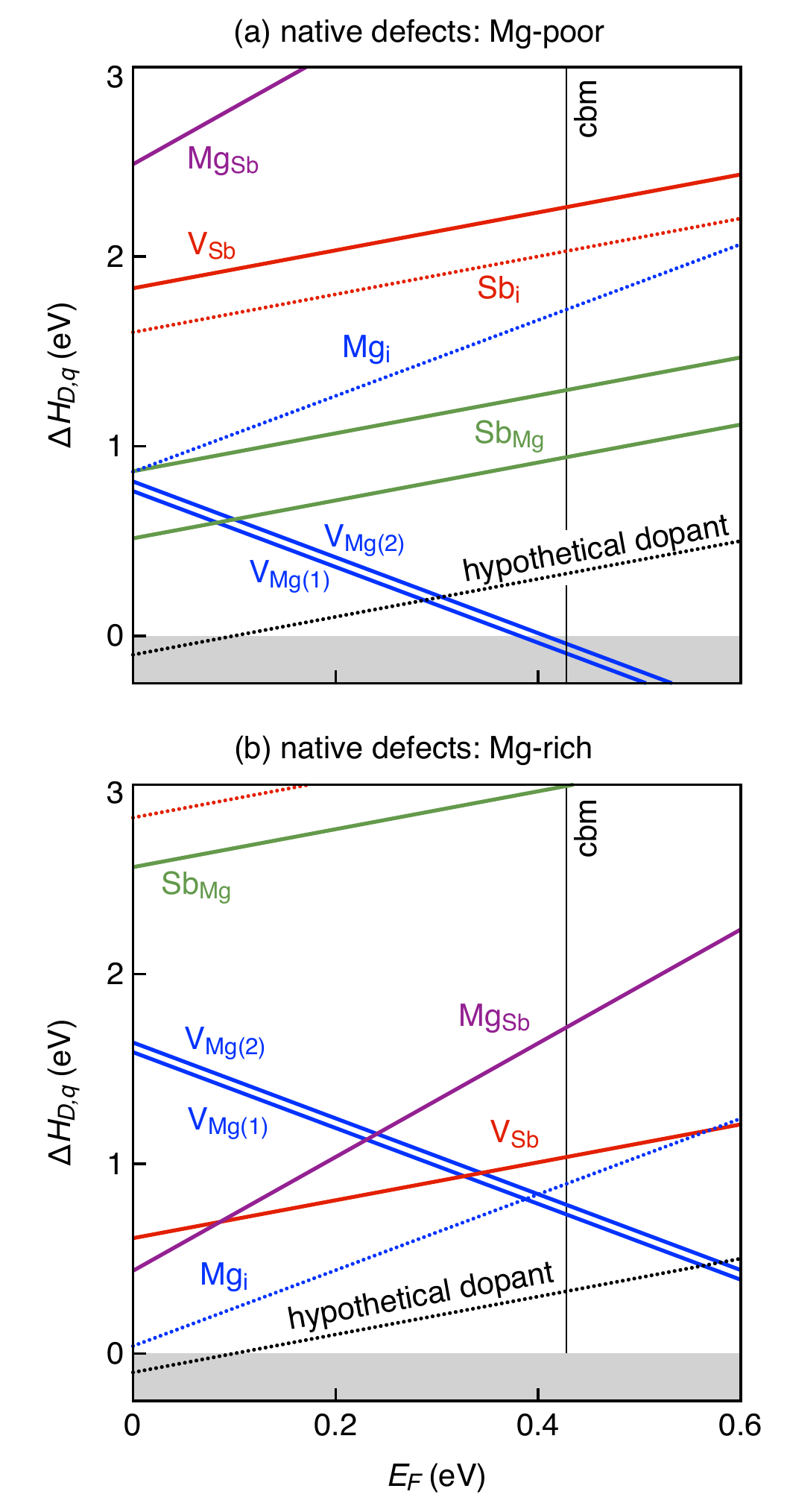}
\caption{\label{native}
Formation energetics of native defects in Mg$_3$Sb$_2$ under (a) Mg-poor, and (b) Mg-rich growth conditions. Subscripts (1) and (2) denote the unique Wyckoff positions of Mg as shown in Figure \ref{structure}. The solid vertical line labeled ``cbm'' is the conduction band minimum. A hypothetical donor dopant is shown with a dotted black line.}
\end{figure}
In previous studies, we have investigated the defect energetics of native point defects \cite{ohno2017} and a suite of candidate $n$-type dopants \cite{Gorai2018}. We briefly revisit these findings in Section \ref{sec:native} to contextualize the results of $n$-type doping with Sc and Y in Section \ref{sec:doping}. 

\subsection{Energetics of Native Defects}\label{sec:native}
%
\begin{figure}[t!]
\centering
\includegraphics[width=0.9\linewidth]{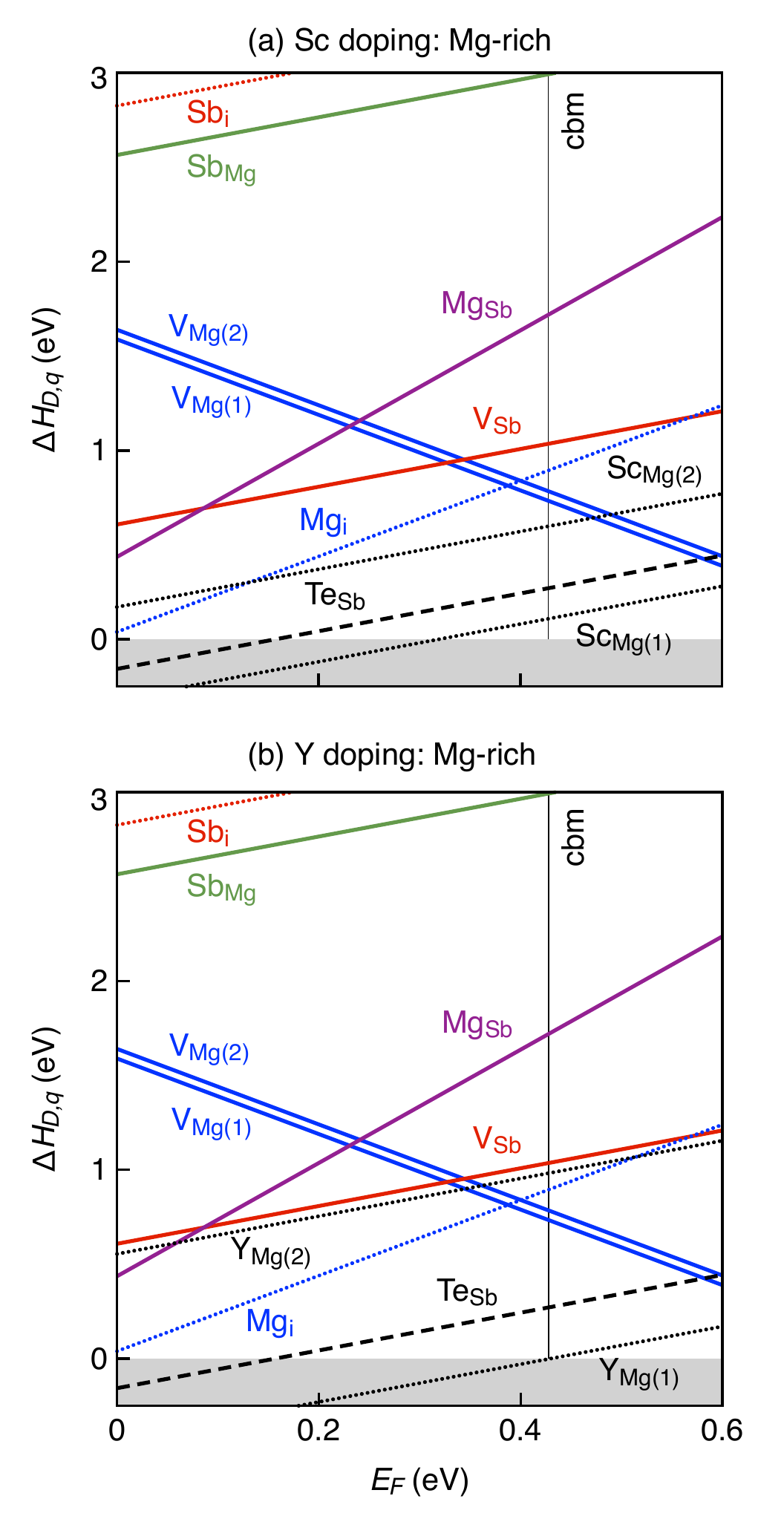}
\caption{\label{scydoping}
Defect energetics of $n$-type doping of Mg$_3$Sb$_2$ by cation substitution under Mg-rich ($\Delta\mu_\mathrm{Mg}$ = 0 eV) and most dopant-rich conditions. The formation energies of native defects and Te doping on Sb site (Te$_\mathrm{Sb}$) are also shown for reference. Doping with group-3 elements: (a) Sc, and (b) Y. The defects corresponding to the extrinsic dopants (Sc$_\mathrm{Mg}$, Y$_\mathrm{Mg}$) are denoted by black dotted lines. Both Sc and Y are found to be more effective $n$-type dopants compared to Te.}
\end{figure}
The formation energetics of native point defects critically influences the dopability of semiconductors, as exemplified in the case of Mg$_3$Sb$_2$. The calculated energetics of native point defects, under Mg-poor and Mg-rich growth conditions, as functions of Fermi energy ($E_F$) in Mg$_3$Sb$_2$ are shown in Figure \ref{native} (adapted from Ref. \citenum{Gorai2018}). Defects plots with positive slopes are donors and negative slopes are acceptors. We found that all native point defects in Mg$_3$Sb$_2$ are shallow; Mg vacancies are the only acceptor defects while all other native defects are donors. More importantly, we found that the growth conditions (Mg-rich \textit{vs.} Mg-poor) have a profound effect on the dopability of Mg$_3$Sb$_2$. Specifically, under Mg-poor conditions, acceptor Mg vacancies are the dominant native defects that hinder $n$-type doping of Mg$_3$Sb$_2$ by compensating or ``killing'' electrons generated by extrinsic donor dopants. This is schematically illustrated in Figure \ref{native}(a), where the electrons generated by the hypothetical donor dopant (dotted black line) is compensated by holes generated by the acceptor Mg vacancies such that the equilibrium $E_\mathrm{F}$ is near mid-gap with low free electron concentration. Commonly, Mg$_3$Sb$_2$ has been synthesized under Mg-poor growth conditions due to the inherently high volatility of Mg; consequently, Mg$_3$Sb$_2$ has been, until recently, synthesized almost exclusively as a $p$-type material \cite{ponnam2013,condron2006,bhardwaj2014}.

In contrast, under Mg-rich growth conditions (Figure \ref{native}b), the formation energy of Mg vacancies are significantly higher and therefore, present in low concentrations. The predominant defects are Mg interstitials that occupy octahedral sites within the [Mg$_2$Sb$_2$]$^{2-}$ layers. Therefore, Mg-rich growth conditions provide an opportunity to extrinsically $n$-type dope  Mg$_3$Sb$_2$; the hypothetical donor dopant (dotted black line) shown in Figure \ref{native}(b) does not suffer from electron compensation.   

\subsection{Energetics of Sc and Y Doping}\label{sec:doping}
%
\begin{figure}
\centering
\includegraphics[width=\linewidth]{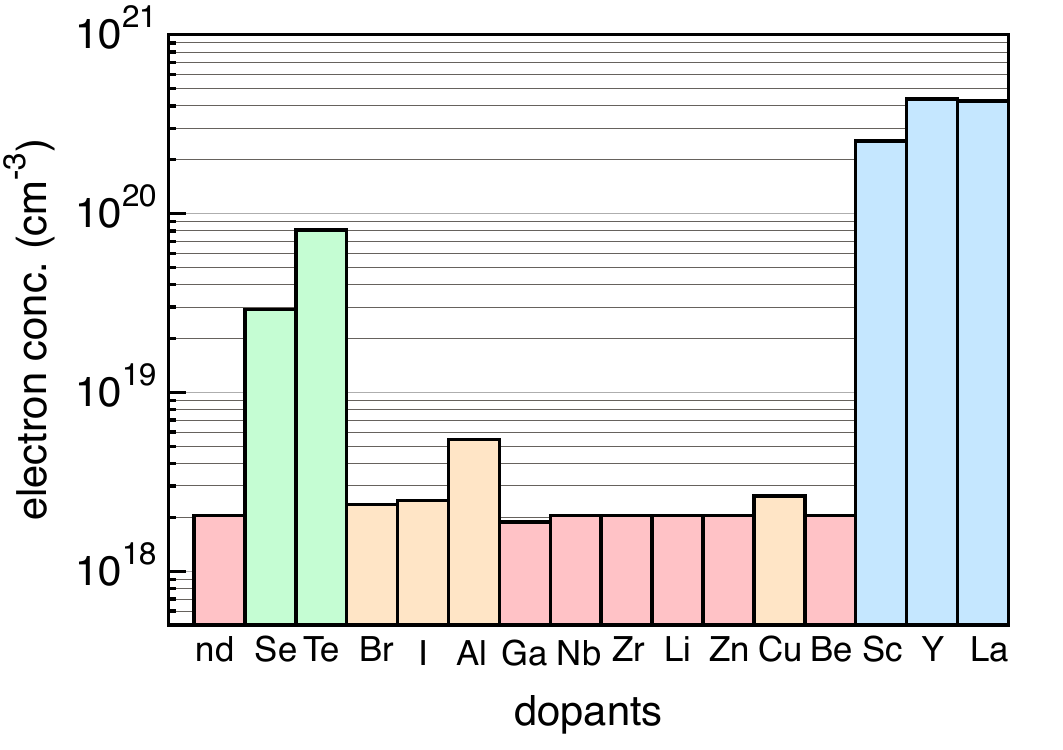}
\caption{\label{dopingcomparison}
Comparison of predicted free electron concentrations (at growth temperature 900 K) for 15 different $n$-type dopants for Mg$_3$Sb$_2$. The free electron concentration in natively self-doped Mg$_3$Sb$_2$ is denoted by ``nd''. The free electron concentrations for all but Sc and Y doping are extracted from the results in Refs. \citenum{Gorai2018} and \citenum{ohno2017}. Group-3 elements (Sc, Y, La) are more effective compared to chalcogens (Se, Te) that are traditionally used for $n$-type doping of Mg$_3$Sb$_2$.}
\end{figure}
The defect energetics of $n$-type doping of Mg$_3$Sb$_2$ with Sc and Y are shown in Figure \ref{scydoping}. Both Sc and Y are effective Mg-site $n$-type dopants due to the low formation energy of substitutional defects Sc$_\mathrm{Mg(1)}$ and Y$_\mathrm{Mg(1)}$. The formation energy of substitutional Te$_\mathrm{Sb}$ in Figure \ref{scydoping} is reproduced from Refs. \citenum{Gorai2018} and \citenum{ohno2017}. We find that $n$-type doping of Mg$_3$Sb$_2$ with Sc and Y is more effective than with chalcogens. The effectiveness is defined as the formation energy of Sc and Y impurities arouns the conduction band minimum. Lower the defect formation energy, higher the solubility of the impurity and consequently, higher the achievable free electron concentration.

As illustrated in Figure \ref{structure}, there are two unique Wyckoff positions of Mg atoms in Mg$_3$Sb$_2$ that are denoted by Mg(1) and Mg(2), where Mg(1) is the Mg atom residing between the [Mg$_2$Sb$_2$]$^{2-}$ slabs. From Figure \ref{scydoping}, we find that Sc and Y preferentially substitutes Mg(1) over Mg(2). The tendency to preferentially substitute Mg(1) is similar to La doping\cite{Gorai2018} and may be attributed to the large ionic size of Sc$^{3+}$ and Y$^{3+}$ ions that can be spatially accommodated betewen the [Mg$_2$Sb$_2$]$^{2-}$ slabs. The substitutional donor defects Sc$_\mathrm{Mg(1)}$ and Y$_\mathrm{Mg(1)}$ have lower formation energies than the donor Te$_\mathrm{Sb}$ (Figure \ref{scydoping}). Doping with Y, in particular, is more effective than Sc and comparable to La doping\cite{Gorai2018}. The predicted free electron concentrations at 900K under Mg-rich and most Y-rich conditions (as allowed by the phase stability of Mg$_3$Sb$_2$) exceeds 10$^{20}$ cm$^{-3}$. 

The high thermoelectric performance of $n$-type Mg$_3$Sb$_2$ is attributed to its highly degenerate conduction band \cite{zhang2017theory} and low lattice thermal conductivity \cite{peng2018}. Heavy doping on the Mg site can have detrimental effects on the intrinsic electron transport properties since the conduction band edge is mainly composed of states derived from Mg($s$) orbitals. The detrimental effects of Mg site doping can be minimized by light doping with an efficitive $n$-type dopant, as experimentally demonstrated for La doping\cite{imasato2018}.

Figure \ref{dopingcomparison} compares the predicted free electron concentrations (at growth temperature 900 K) for a suite of 15 plausible $n$-type dopants in Mg$_3$Sb$_2$. It is evident from Figure \ref{dopingcomparison} that group-3 elements (Sc, Y, La) are far more effective $n$-type dopants compared to chalcogens (Se, Te). With group 3 elements, free electron concentrations exceeding 10$^{20}$ cm$^{-3}$ can be achieved. Recent experimental confirmation\cite{imasato2018} of effective $n$-type doping with La provides further confidence in our prediction about effective $n$-type doping with group-3 elements.

\section{Conclusions}
In this work, we have computationally assessed the effectiveness of $n$-type doping of Mg$_3$Sb$_2$ with group-3 elements such as Sc and Y. We find that both Sc and Y are more effective $n$-type dopants than chalcogens (Se, Te). The predicted doping behaviors of Sc and Y are similar to La; Y is more effective than Sc with predicted free electron concentrations comparable to those achieved by La doping. The higher doping effectiveness of Sc and Y offers greater tunability of the electronic properties in Mg$_3$Sb$_2$-based thermoelectric materials. Also, the relative abundance of group-3 elements (Sc, Y, La) compared to Te offers at additional incentive for considering these dopants. Following the successful experimental confirmation\cite{imasato2018} of our prediction about effective La doping\cite{Gorai2018} of Mg$_3$Sb$_2$-based materials, we call upon the experimentalists to confirm the new predictions presented in this work.\\

\section*{Acknowledgements}
We acknowledge support from NSF DMR program, grant no. 1729594. The research was performed using computational resources sponsored by the Department of Energy's Office of Energy Efficiency and Renewable Energy and located at the National Renewable Energy Laboratory (NREL).

%

\end{document}